\documentclass{aa}  
\usepackage{graphicx}
\usepackage{txfonts}
\begin{document}
%

\title{Biomarkers in disk-averaged near-UV to near-IR Earth spectra using Earthshine observations
\thanks{Based on NTT/EMMI observations collected at ESO-La Silla, Chile, programs 72.C-0125A, 73.C-0010A and DDT 275.C-5015}
   }

   \subtitle{}

   \author{S. Hamdani
          \inst{1}
          \and
          L. Arnold \inst{1}
          \and
          C. Foellmi \inst{2}
          \and
          J. Berthier \inst{3}
          \and
          M. Billeres \inst{2}
          \and
          D. Briot \inst{4}
          \and
          P. Fran\c cois \inst{4}
          \and
          P. Riaud \inst{5}
          \and
          J. Schneider \inst{4}
          }

   \offprints{hamdani@obs-hp.fr;\ arnold@obs-hp.fr. \\S. Hamdani is now at EGO/VIRGO - Via E. Amaldi 56021 S. Stefano a Macerata - Cascina (PI), Italia}

   \institute{Observatoire de Haute Provence - CNRS, 04870 Saint Michel l'Observatoire, France\\
              \email{hamdani@obs-hp.fr, arnold@obs-hp.fr}
         \and
             ESO, Casilla 19001, Santiago 19, Chile; now at Laboratoire d'Astrophysique, Observatoire de Grenoble, BP 53, 38041 Grenoble Cedex 9, France\\
             \email{cfoellmi@eso.org, mbillere@eso.org}
          \and
             IMCCE, Observatoire de Paris, 77 Avenue Denfert-Rochereau, 75014 Paris, France\\
             \email{berthier@imcce.fr}    
          \and
             Observatoire de Paris-Meudon, 5 place Jules Janssen 92195 Meudon, France\\
             \email{danielle.briot@obspm.fr, patrick.francois@obspm.fr, jean.schneider@obspm.fr} 
           \and
             	Institut d'Astrophysique et de G\'eophysique de Li\`ege, Universit\'e de Li\`ege, All\'ee du 6 Ao\^ut, 4000, Sart-Tilman, Belgium\\
             \email{riaud@mesiog.obspm.fr}             
             }

   \date{Received 16 February 2006; accepted 11 August 2006}

 
  \abstract
{The detection of exolife is one of the goals of very ambitious future space missions or extremely large ground-based telescopes that aim to take direct images of Earth-like planets. While associations of simple molecules present in the planet's \textit{atmosphere} ($O_2$, $O_3$, $CO_2$ etc.) have been identified as possible global biomarkers, we analyse here the detectability of vegetation on a global scale on Earth's surface. 
}
{Considering its specific reflectance spectrum showing a sharp edge around 700 nm, vegetation can be considered as a potential global biomarker. This work, based on observational data, aims to characterise and to quantify this signature in the disk-averaged Earth's spectrum.}
{Earthshine spectra have been used to test the detectability of the "Vegetation Red Edge" (VRE) in the Earth spectrum. We obtained reflectance spectra from near UV (320 nm) to near IR (1020 nm) for different Earth phases (continents or oceans seen from the Moon) with EMMI on the NTT at ESO/La Silla, Chile. We accurately correct the sky background and take into account the phase-dependent colour of the Moon. VRE measurements require a correction of the ozone Chappuis absorption band and Rayleigh plus aerosol scattering.}
{The near-UV spectrum shows a dark Earth below 350 nm due to the ozone absorption. The Vegetation Red Edge is observed when forests are present ($4.0\%$ for Africa and Europe), and is lower when clouds and oceans are mainly visible ($1.3\%$ for the Pacific Ocean). Errors are typically $\pm0.5$, and $\pm1.5$ in the worst case. We discuss the different sources of errors and bias and suggest possible improvements.}
{We showed that measuring the VRE or an analog on an Earth-like planet remains very difficult (photometric relative accuracy of 1\% or better). It remains a small feature compared to atmospheric absorption lines. A direct monitoring from space of the global (disk-averaged) Earth's spectrum would provide the best VRE follow-up.
   }


\maketitle
%

\section{Introduction}
Future space missions like Darwin/TPF will provide the first images and low-resolution spectra of Earth-like extrasolar planets, with the final goal to extract a ''biosignature'' from the spectra. The TPF-C NASA mission will observe exo-terrestrial planets around the nearby stars with a large off-axis telescope equipped with a visible coronagraphic device. To prove the interest of the visible part of the spectrum in spite of the huge contrast ($10^{10}$) compared to thermal infrared ($10^7$), we study the presence of biomarker features.
The spectrum of the light reflected by a planet, when normalized to the parent star spectrum, gives the planet reflectance spectrum, revealing planet atmospheric and ground colour, when the latter is visible through a partially transparent atmosphere. The planet will remain unresolved and its spectrum will be spatially integrated (i.e. disk-averaged) at the observed orbital phase of the planet. In this work, we estimate how the Earth's spectrum would look like from $\sim$ 10 pc. This can be done - in principle at least - by integrating spatially-resolved spectra from low-orbit satellites. It can also be done from a space probe traveling into the  Solar System and looking back the Earth, as Mars Express did in 2003\footnote{http://www.esa.int/esaCP/Pr\_ 44\_ 2003\_ p\_ EN.html}. 
An alternative method to obtain the Earth's disk-averaged spectrum consists of taking a spectrum of the Moon Earthshine, i.e. Earth light backscattered by the non-sunlit Moon. A spectrum of the Moon Earthshine directly gives a disk-averaged spectrum of the Earth at a given phase as seen from the Moon (since the Moon surface roughness "washes out" any spatial information on the Earth's colour). 

Biosignatures are of two types: out-of-equilibrium molecules in the planet atmosphere (like oxygen and ozone) or ground colours characteristic of biological complex molecules (like pigments in vegetation). The visible and near infra-red Earthshine spectra published to date (\cite{arnold02,woolf02,seager05,montanes05}) clearly show the atmospheric signatures and, at least, tentative signs of ground vegetation which thus appears as an interesting potential global biomarker. 
Vegetation indeed has a high reflectivity in the near-IR, higher than in the visible by a factor of $\approx5$ (\cite{clark99}). This produces a sharp edge around $\approx 700\rm nm$, the so-called Vegetation Red Edge (VRE). An Earth disk-averaged reflectance spectrum can thus rise by a significant fraction around this wavelength if  vegetation is in view from the Moon when the Earthshine is observed. Arcichovsky suggested in 1912 to look for chlorophyll absorption in the Earthshine spectrum, to calibrate chlorophyll in the spectrum of other planets (\cite{arci1912}). This approach re-emerged again only in 1999 with preliminaryt tests at ESO by several among us (P.F., J.S. and D.B.) and at OHP (\cite{arnold02}). The position of the VRE may not lie at 725 nm on an exo-Earth. Its possible confusion with mineral spectral features is discussed by Schneider (2004) and Seager et al. (2005).

The red side [600:1000 nm] of the Earth reflectance spectrum shows the presence of $O_2$ and $H_2O$ absorption bands and of the VRE, while the blue side [320:600 nm] clearly shows the Huggins and Chappuis ozone ($O_3$) absorption bands. The higher reflectance in the blue shows that our planet is blue due to Rayleigh scattering in the atmosphere, as detected by Tikhoff (1914) and Very (1915), and confirmed later with accurate Earthshine photometry by Danjon (1936).
 
The present paper presents new observations and emphasizes the difficulties of data reduction  and describes how it has been improved with respect to our first work (\cite{arnold02}). In Section \ref{background_sub}, we explain how we adapt Qiu's method (\cite{qiu03}) to carefully subtract the sky background from Earthshine spectra. Section \ref{sec:moon} describes how we correct the phase-dependent colour effect of the Moon, following a new processing of Lane \& Irvine (1973) data. This correction was not done in Arnold et al. (2002). 
Section \ref{discussion} presents the first - as fas as we know - near-UV integrated Earth spectra obtained from Earthshine observations that reveal a dark Earth below
350 nm due to the strong $O_3$ absorption (ozone Huggins absorption bands). While this work presents smaller VRE than Arnold et al. (2002) and Woolf et al. (2002), it 
nevertheless shows different values depending whether ocean or land are seen from the Moon at the epoch of observations.

\section{Observations}\label{sec:data}
The observations were taken at the NTT/La Silla telescope (3.5m) on July 24th and September 18th 2004 for the descending Moon and May 31st and June 2nd 2005 for the ascending Moon (Table \ref{data}). The Earthshine spectrum for 05-31-2005 was taken just before the last quarter of the Moon and has a very low contrast compared to the sky background. The spectra were obtained with the EMMI spectrograph in the medium dispersion mode in the blue (BLMD mode) and in low dispersion in the red (RILD mode), enabling us to record spectra from 320 to 1020 nm, with a 20 nm gap around 520 nm. The spectral resolution is R$\cong$450 in the blue and
 R$\cong$250 in the red.
To record both Earthshine (hereafter ES) and sky background spectra simultaneously, EMMI's long slit was oriented East-West on the lunar limb (\cite{arnold02}). ES exposures are bracketed by at least two exposures of the bright Moonshine (hereafter MS) with the slit oriented North-South. The length of the slit (6-arcmin in blue and 8-arcmin in red modes) allows us to sample the Moon spectrum over a large lunar region, giving a correct mean of the Moon spectrum.
MS spectra are recorded through a neutral density filter in the blue arm and with a diaphragm in the red arm (unfortunately the diaphragm could not be placed exactly in a pupil plane, resulting in strong vignetting - no neutral density filter was available in the red).

\begin{table*}
\caption{Dates of observations}             
\label{data}      
\centering          
\begin{tabular}{llccc}     
\hline\hline       
Date  &  Hour    &   Exposure Time & Exposure Time & MS phase angle \\
      & (UT)   & (blue)  & (red)\\ 
\hline                    
05/24/2004   & 23h16(blue)               & 2x180sec.    & - & -117\char23 \\
09/18/2004   & 0h15(red)                 & -            & 2x100sec. & -137\char23 \\
05/31/2005   & 9h30(red) and 10h01(blue) & 2x250sec.    & 2x120sec. & 101\char23 \\
06/02/2005   & 9h25(red) and 9h50(blue)  & 2x250sec.    & 2x120sec. & 126\char23 \\\hline                 
\end{tabular}
\end{table*}

\section{Spectra processing method}

Data reduction is done with dedicated IDL\texttrademark routines. All images are processed for cosmic rays, bias, dark, flat corrections (dome-flat) and
distortion. Some of the MS blue spectra have been obtained through a neutral density filter (ND=0.3) and these files have been corrected from its not-perfectly-flat spectral profile. The Earthshine spectrum needs to be corrected for the bright sky background (comparable to the Earthshine flux), the Earth atmosphere transmittance and phase-dependent colour effects of the Moon reflectance.

\subsection{Sky background subtraction}
\label{background_sub}
\begin{figure}
\scalebox{0.75}{\includegraphics[viewport=20 10 110 175]{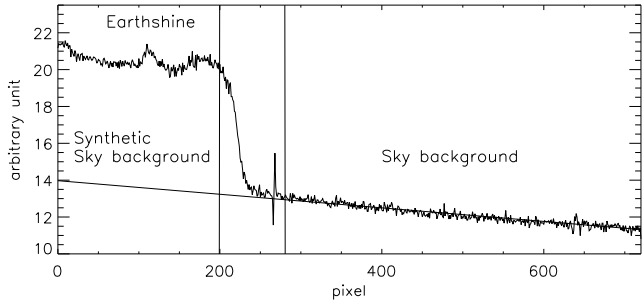}} 
\caption{Flux at a given wavelength along the slit placed perpendicular to the edge of the Moon to record both Earthshine and sky background simultaneously. The linear fit based on the sky background is extrapolated through the Moon surface to estimate the synthetic sky background to be removed from the Earthshine data. The transition zone between ES and background (pixel 200 to 280) is not used in the data processing.}
\label{fondciel}
\end{figure}

\begin{figure}
\scalebox{0.75}{\includegraphics[viewport=20 10 110 175]{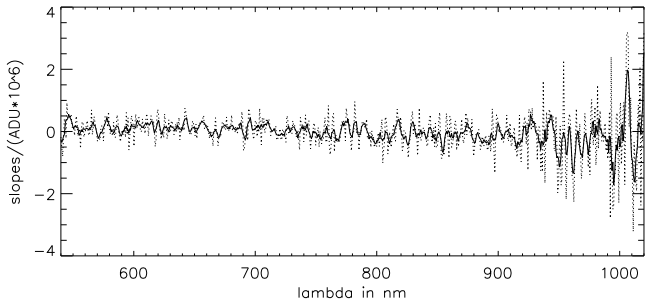}} 
\caption{The slopes obtained at each wavelength from the sky background (Fig. \ref{fondciel}) are normalized to the mean flux of the recorded sky background (dashed line) and smoothed (solid line).}
\label{fitslopes}
\end{figure}

A fraction of the sunlit Moon flux is scattered by the atmosphere and produces a bright background around the Moon, decreasing with the distance to the sunlit crescent. Both sky and Earthshine must be corrected from this ''pollution''. We consider that this noise decreases linearly with distance (\cite{qiu03}). The sky background spectrum recorded near the ES spectrum can thus be linearly extrapolated to find its value over the Moon surface. We first make a linear extrapolation at each wavelength to construct what we call a synthetic sky background to be subtracted from the Earthshine (Fig. \ref{fondciel}). The figure is in agreement with the linear assumption mentioned above. If we plot the slopes obtained at each wavelength for the linear extrapolation, we notice strong variations between adjacent wavelengths. These variations are proportional to the flux at each wavelength, but are also due to the lower S/N in the absorption lines. To reduce the noise on the strongest telluric lines, the slopes of sky background are normalized to the mean flux at each wavelength and smoothed (boxcar average of 4 to 20 pixels width depending on the data, Fig. \ref{fitslopes}). The final synthetic sky background spectrum is obtained by multiplying the previous smoothed spectrum by the sky background mean flux at each wavelength.
Once the background is corrected, each image is binned into an improved S/N ratio spectrum.

\subsection{The Earth reflectance ER$(\lambda)$} \label{reflectance}
The Earth reflectance ER is given by the ratio of the Earthshine spectrum divided by the sunlit Moon spectrum (\cite{arnold02})
\begin{equation}
  ER(\lambda)=\frac{ES(\lambda)}{MS(\lambda)}r(\lambda).
 \label{albedo}
\end{equation}
Note that the ES spectrum contains the signature of roughly three paths through the Earth atmosphere: Two paths results from direct Sun light that is afterward scattered toward the Moon and a third one from the light coming back from the ES. MS results from only one path. Therefore the division in Eq.\ref{albedo} exactly corrects one path through the atmosphere, which is identical for both ES and MS spectra. Thus ER contains the absorption signature of roughly two paths through the atmosphere (see \cite{arnold02} for details).

Eq.\ref{albedo} assumes that ES($\lambda$) and MS($\lambda$) are recorded simultaneously, i.e. at the same air-mass. In practice, each ES exposure is bracketed by two MS exposures. To estimate the MS spectrum at the epoch of the ES exposure, the MS is obtained from the average of the two MS spectra, weighted proportionally to the time elapsed before and after the ES exposure.
For the blue spectra of the MS, we also need to correct for the effect of the neutral density filter that does not have the same absorption at each wavelength.
The $r(\lambda)$ is a geometrical chromatic factor that takes into account the geometrical positions of the Sun, the Earth and the Moon. It is described in the next section. 

A photometrically calibrated Earth reflectance spectrum can be obtained in principle by following the procedure used for broad-band albedo measurements (\cite{qiu03}). This requires us to measure spectra or brightness only in calibrated areas to properly take into account the Moon phase function. Our data do not allow us to calibrate all fluxes, so our Earth reflectance spectra are not absolutely calibrated.

\subsection{The Moon phase colour effect}\label{sec:moon}
Lane \& Irvine (1973) described the chromatic dependence of the integrated Moonshine versus the lunar phase. Their narrow band photometry showed a $\approx$30\% excess of red light at 1050 nm with respect to the flux at 350 nm at phase=120\char23. 
Lane \& Irvine plotted MS fluxes versus phase angles for different wavelengths. They identified a linear domain for low phase angles ($|\rm{phase}|<40^{\circ}$) and applied a cubic fit for $|\rm{phase}|>40^{\circ}$. Their cubic fit is strongly constrained by a very few data points at high phase angles (around 120\char23), which probably biases their fit. They also observed a difference of brightness between the waxing and waning Moon observed at exactly opposite phase angles. This is due to the different proportion of visible craters and Mares (\cite{rougier33,russell16})  with respect to the Moon phase. But Lane \& Irvine report that the difference they measured seems too large. They processed only their second set of data, with the majority of measurements at positive phase angles. This slightly biased their results toward positive phase angles, but their data were nevertheless consistent with Rougier's data (1933), the most accurate at that time. 

Applying the Lane \& Irvine correction to our data leads to a positive slope of the ER spectrum in the [400:520 nm] domain. This is inconsistent with the dominant Rayleigh scattering which gives a negative slope in the blue for a cloud-free Earth, or, at worst, a zero slope for a white Earth fully covered by clouds. Paillet \& Selsis (2005, private communication) confirmed the permanent negative slope in the blue with simulated Earth reflectance spectra obtained with a radiative transfer code. Therefore, the Lane \& Irvine original fit leads to a clear over-correction of the colour phase effect.

\begin{figure}
\scalebox{0.75}{\includegraphics[viewport=20 10 110 175]{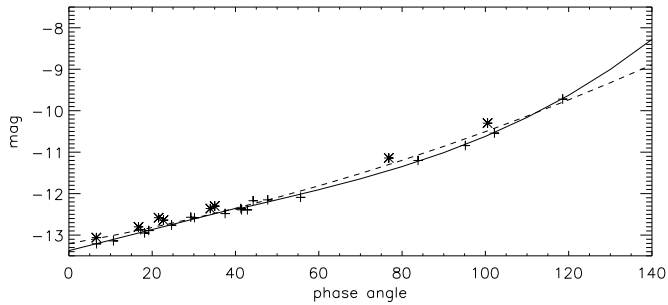}} 
\caption{Lane \& Irvine data at 1063.5 nm: stars and crosses correspond to data from 1964 and 1965, respectively. The solid line is the original Lane \& Irvine linear ($\le$40\char23\hspace{1pt}) and cubic ($>$40\char23\hspace{1pt}) fit based on the 1965 set of data only. The dash line is the second order fit based on all 
data and used in this work to correct the phase-dependent colour of the Moon.}
\label{fit_lane}
\end{figure}

\begin{figure}
\scalebox{0.75}{\includegraphics[viewport=20 5 110 175]{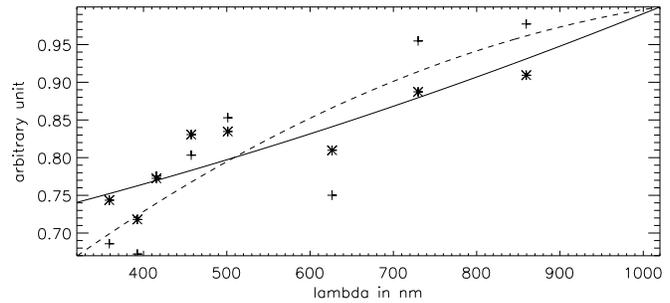}} 
\caption{Data extrapolated at a Moon phase of -137\char23. The crosses are determined with a linear-cubic fit and the star with our second order fit. The data are normalised at 1020 nm for easier comparison. A second order fit is used to define the colour correction (plain line for the star and dotted line for the crosses).}
\label{fit_correction}
\end{figure}

We therefore reconsidered all the Lane \& Irvine data and fit all points with a second-order curve over the full range of phase angles, without considering two distinct regimes (i.e. linear + cubic). An example of the fit is shown in Fig. \ref{fit_lane}. We processed each wavelength and extrapolated the correction at the phase angles of our observations. Another second-order interpolation through the nine obtained values (the nine Lane \& Irvine photometric filters) gives the $r(\lambda)$ factor needed to correct the reflectance spectra given by Eq. \ref{albedo} (Fig. \ref{fit_correction}).

We have to observe two regions of the Moon, the ES and the MS. While MS is observed at high phase angles (i.e. a narrow crescent), ES is observed at 0\char23\hspace{1pt} phase angle because the Moon source of light and the observer are both on Earth. This difference in phase angles induces a colour difference between ES and MS that must be taken into account in the data reduction. This is done with the $r(\lambda)$ factor (Eq. \ref{albedo}) which allows one to calculate the MS spectrum at 0\char23\hspace{1pt} phase angle, i.e. the same as for ES.

We also compared each individual ES spectrum line along the slit going from the edge to the center of the Moon and detected a smooth chromatic variation along the slit, resulting in a redder Moon at the center than at the edge of the lunar disk.  
This residual chromatic variation is due to i) spectral changes due to the soil nature (crater versus Mare), and ii) a chromatic dependence of the spectra along the spectrograph slit, due to the Bidirectional Reflectance Distribution Function (BRDF) of the soils sampled along the slit. Most of the phase-dependent colour effect is canceled by the Lane \& Irvine correction, but it provides only a mean correction based on observations of the integrated illuminated Moon at a given phase, while here we speak about differences from point to point on the lunar surface. In our spectra, the maximum increase in the red between 320 and 1020 nm is of the order of 3\%. The measurement of this residual colour effect however remains too uncertain. It seems smooth and minor so we have neglected it in our data reduction, although we estimate an induced bias of +0.15 on the VRE.

\subsection{The VRE measurement method}
\label{vre_method}
\begin{figure}
\scalebox{0.75}{\includegraphics[viewport=20 10 110 175]{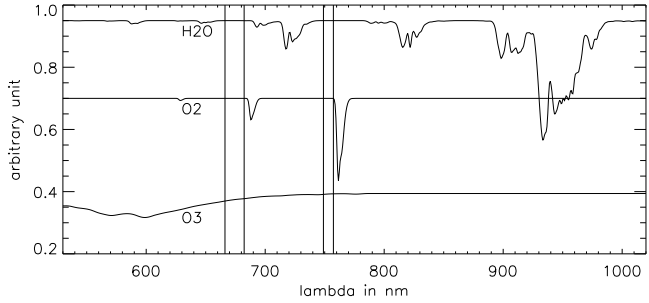}}
\caption{Spectra of the main atmospheric absorbing components (shifted vertically for clarity). Vertical lines indicate the two spectral bands used to calculate the VRE, [667:682 nm] and [745:752 nm], and show that only $O_3$ intersects the domain where the VRE is measured.}
\label{atmosphere}
\end{figure}

To detect the vegetation signature, i.e. a signature of the ground, in principle we need to extract the Earth surface reflectance from the ER, which contains the signatures of the Earth atmosphere. To remove the atmospheric signatures, ER can be divided by the atmosphere transmittance AT($\lambda$) that can be obtained from the ratio of two spectra of a calibration star (or MS) taken at two different air-masses (\cite{arnold02}). It can also be built from spectral databases and adjusted to fit the observed spectrum (\cite{woolf02}). In this work, we only correct the ER spectrum for the $O_3$ absorption because, among the main atmospheric absorption lines ($O_2$, $O_3$ and $H_2O$), only the broad Chappuis band extends up to $\approx700$ nm and thus can potentially induce a bias in the red part of the spectrum where the VRE is measured (Fig. \ref{atmosphere}).

\begin{figure}
\scalebox{0.75}{\includegraphics[viewport=20 10 110 175]{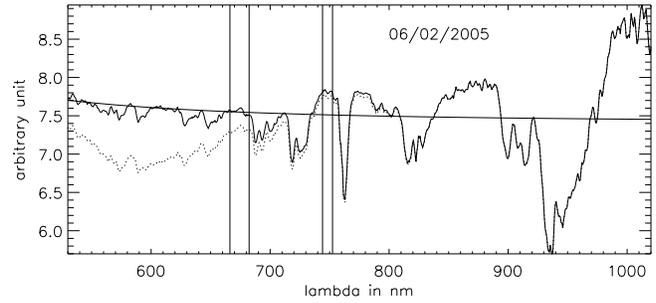}}
\caption{ER($\lambda$) spectrum (dotted line) and corrected for the $O_3$ Chappuis absorption band (solid line). The smooth solid line is a fitted Rayleigh-aerosol scattering model. Vertical lines define the two spectral bands used to calculate the VRE.}
\label{fit}
\end{figure}

The ER($\lambda$) spectra are fitted with $O_3$ laboratory measured spectra (\cite{voigt01}) convolved to reproduce EMMI's spectral resolution. Removing of the Chappuis band requires great care: we observed that the band can indeed easily be slightly over-corrected, leading to an apparently smooth Rayleigh scattering but an underestimated VRE.
In the [520:670 nm] domain, the ER spectrum is dominated by the absorption of $O_3$, Rayleigh and aerosol scattering. To remove the $O_3$ Chappuis band, we fit a Rayleigh-aerosol function based on continuum parts of the spectrum for different concentration of $O_3$. We then fit a second-order polynomial on the root mean squares of the difference between the fit and the spectrum for each $O_3$ concentration to find the best fit, i.e. the minimum root mean square and its corresponding $O_3$ concentration. The horizontal RMS of this second fit gives an estimate of the ozone column.
We define the Rayleigh-aerosol function as the sum of a Rayleigh scattering function $a + b \lambda^{-4}$ plus an aerosol scattering contribution $c \lambda^{-1.5}$. To reduce the number of parameters, we consider that the aerosol and the Rayleigh contribution are equal at $\lambda_e=700$ nm (\cite{lena96}), so the scattering law is
\begin{equation}
\rm{scat}({\lambda})=a+b\ \big(\frac{1}{\lambda^4}+\frac{1}{\lambda_e^{2.5}\lambda^{1.5}}\big).
\end{equation}
Once the Chappuis band is corrected, the spectrum is again fitted with a Rayleigh-aerosol function in continuum domains and normalised to that function to obtain a spectrum where the vegetation signature can be quantified, although the spectrum is not corrected for $O_2$ and $H_2O$. The VRE is measured on this corrected spectrum (Fig. \ref{fit}) following Eq.\ref{eqVRE}: 
\begin{equation}
VRE=\frac{r_I-r_R}{r_R}
\label{eqVRE}
\end{equation}
where $r_R$ and $r_I$ are the mean reflectance in the [667:682 nm] and [745:752 nm] spectral bands.

\begin{figure}
\scalebox{0.75}{\includegraphics[viewport=20 10 110 175]{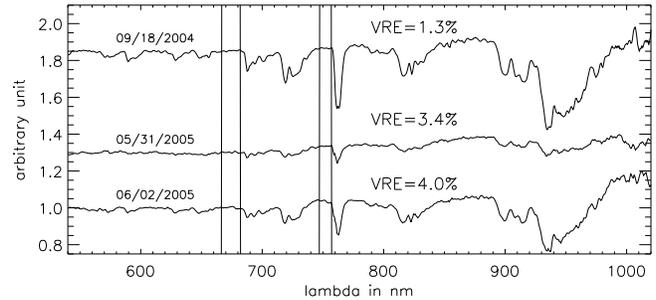}} 
\caption{Resulting ER($\lambda$) corrected for $O_3$ Chappuis absorption, Rayleigh and aerosol scattering, thus ready for the VRE measurement. Vertical lines define the two spectral bands used to calculate the VRE. The plots have been shifted vertically for clarity. }
\label{VRE}
\end{figure}

\section{Results and discussion}
\label{discussion}
\subsection{Qualitative analysis}
\begin{figure*}[!]
\scalebox{0.95}{\includegraphics[viewport=25 10 128 175]{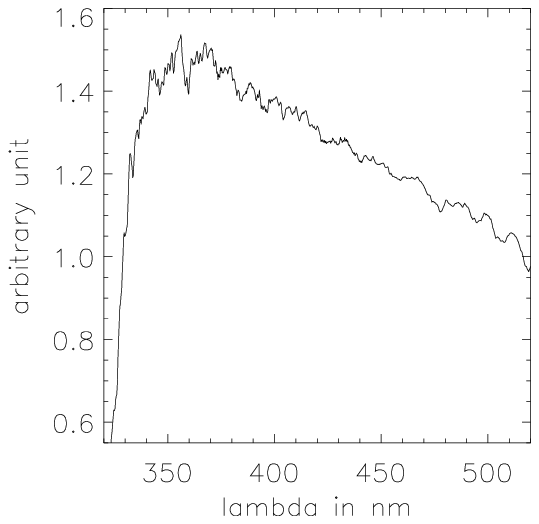}}
\scalebox{0.95}{\includegraphics[viewport=3 10 343 175]{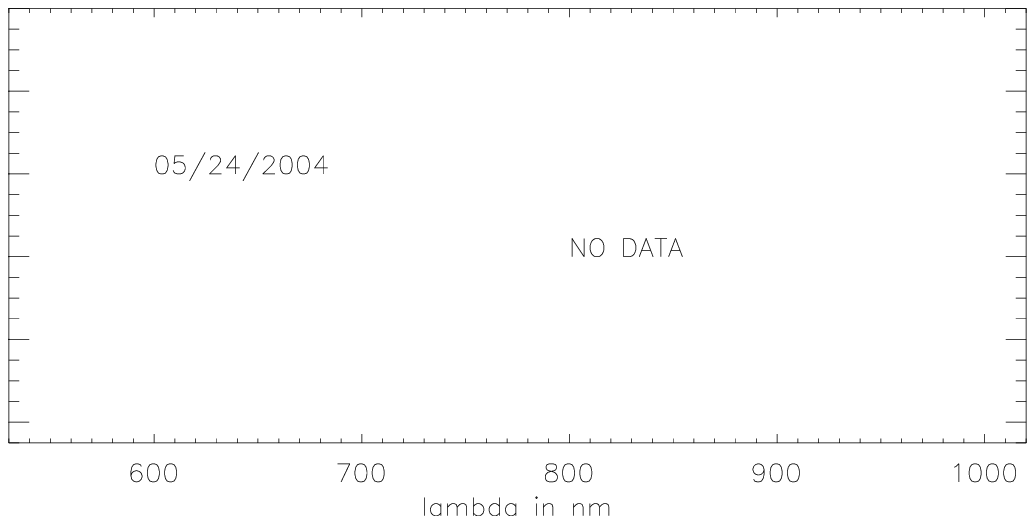}}
\scalebox{0.33}{\includegraphics{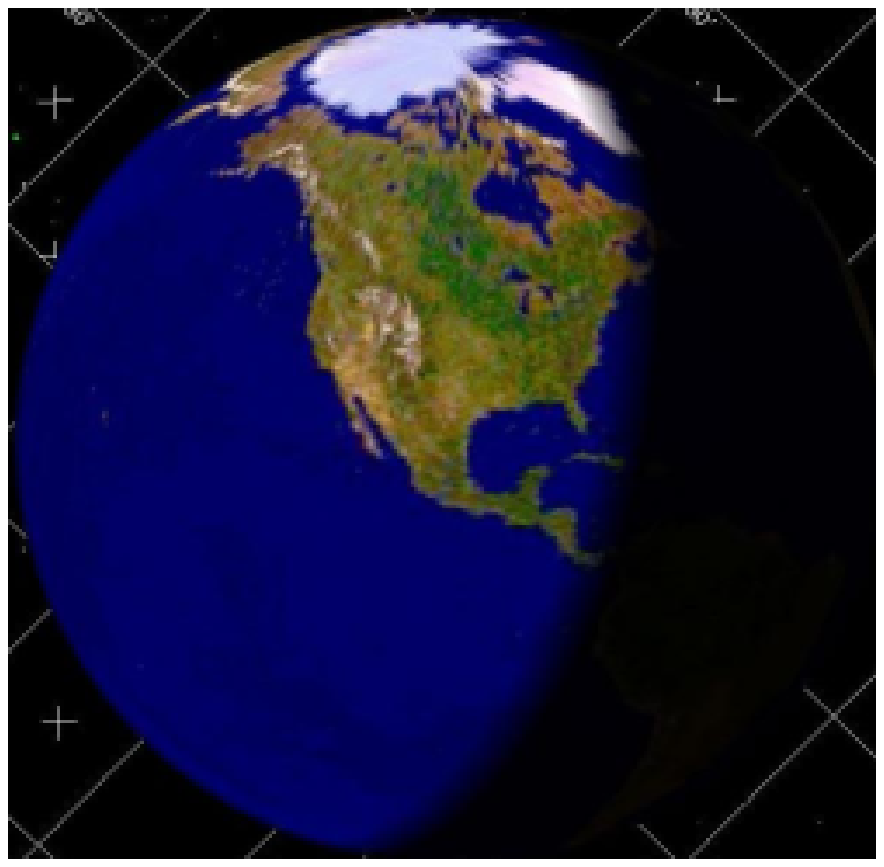}}

\scalebox{0.95}{\includegraphics[viewport=25 10 128 175]{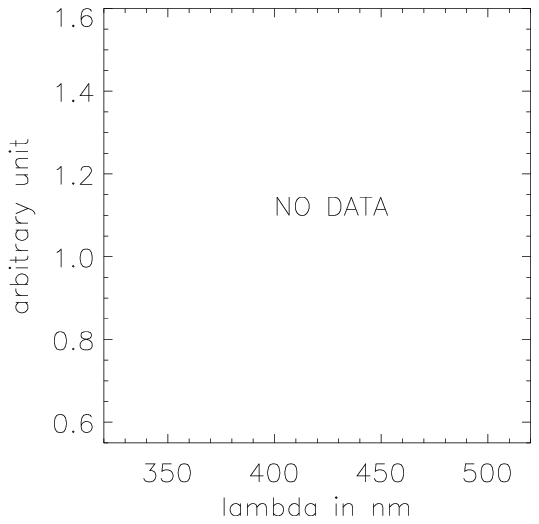}}
\scalebox{0.95}{\includegraphics[viewport=3 10 343 175]{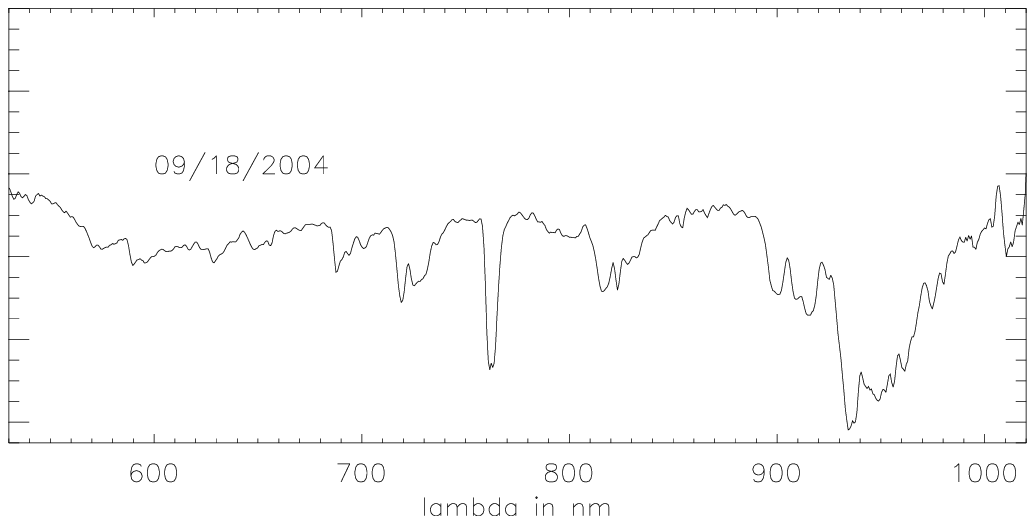}}
\scalebox{0.33}{\includegraphics{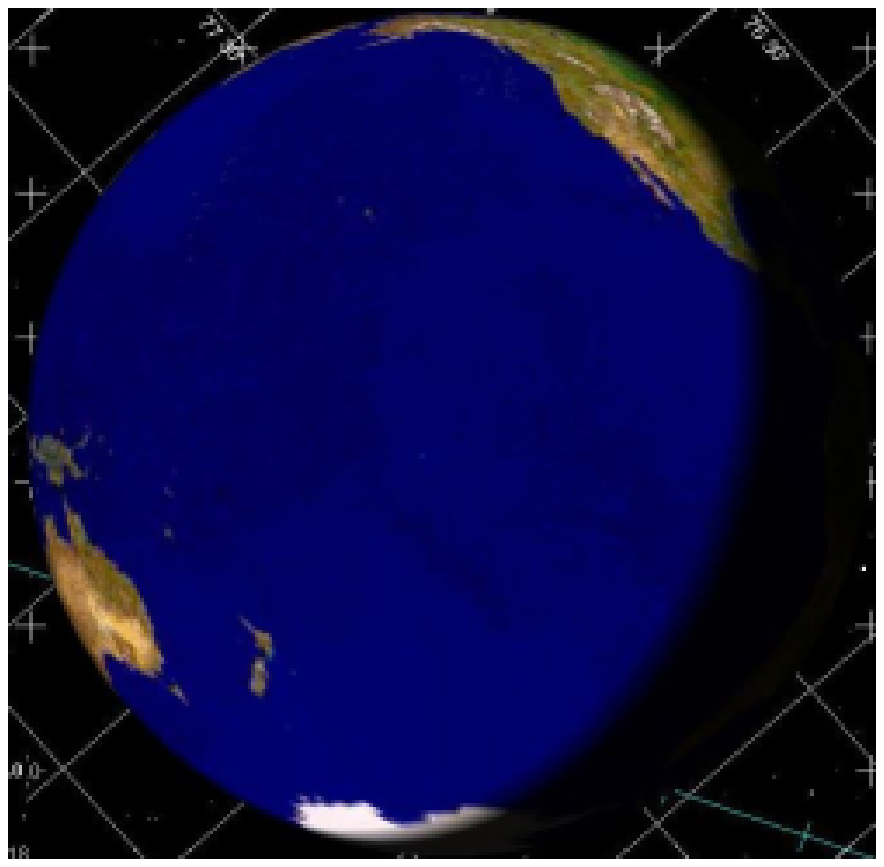}}

\scalebox{0.95}{\includegraphics[viewport=25 10 128 200]{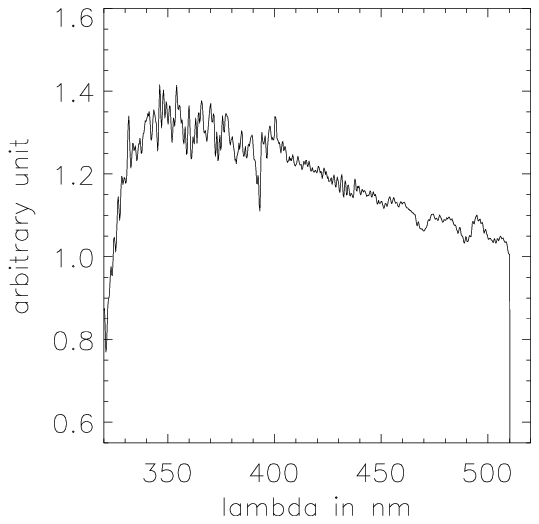}}
\scalebox{0.95}{\includegraphics[viewport=3 10 343 175]{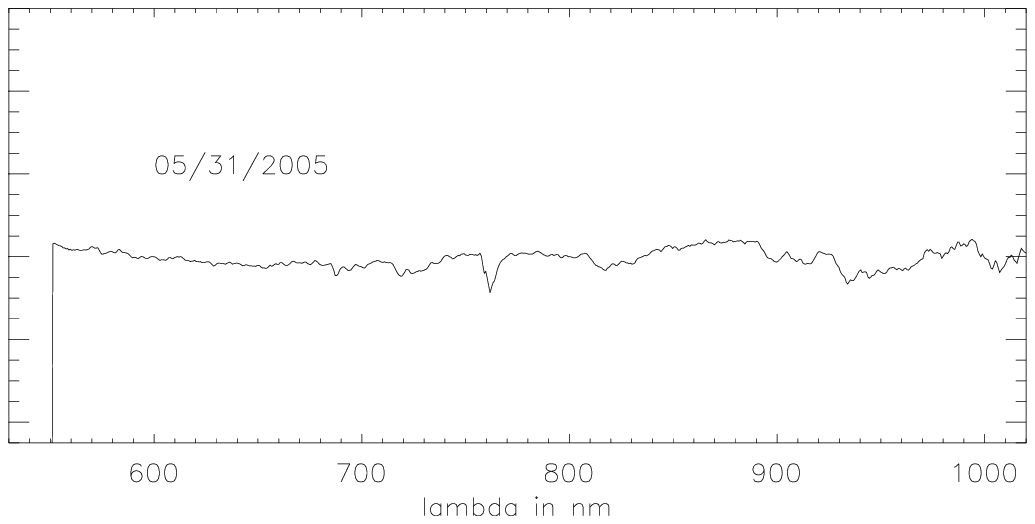}}
\scalebox{0.33}{\includegraphics{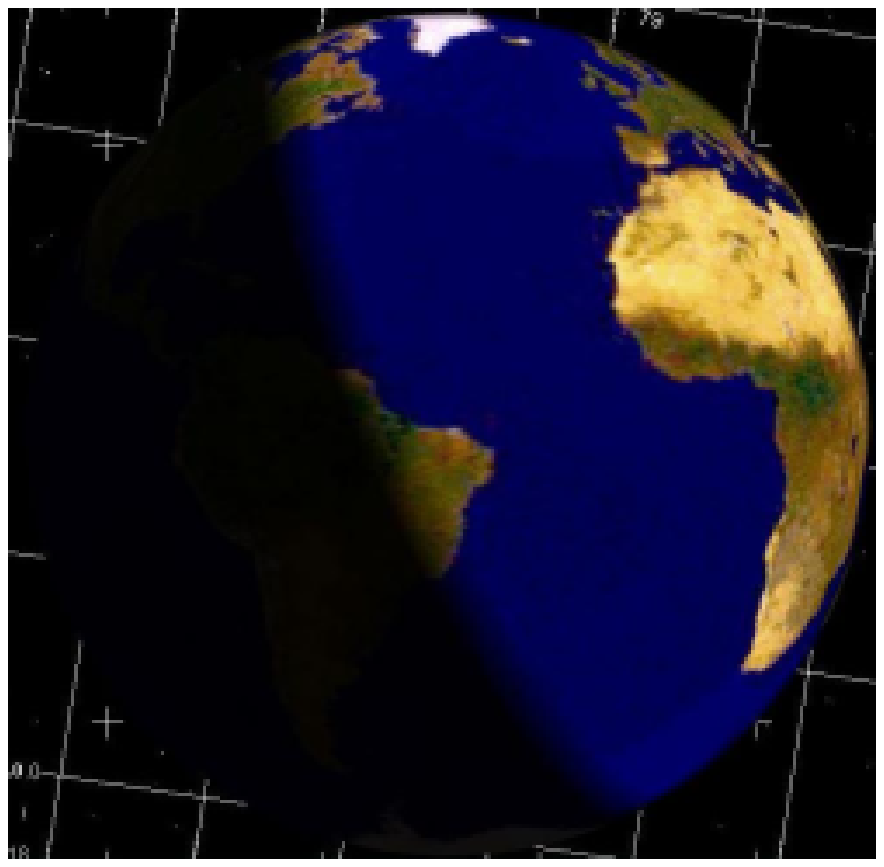}}

\scalebox{0.95}{\includegraphics[viewport=25 10 128 175]{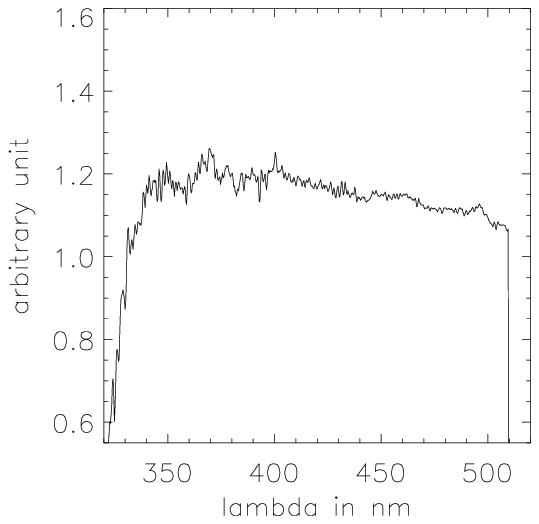}}
\scalebox{0.95}{\includegraphics[viewport=3 10 343 175]{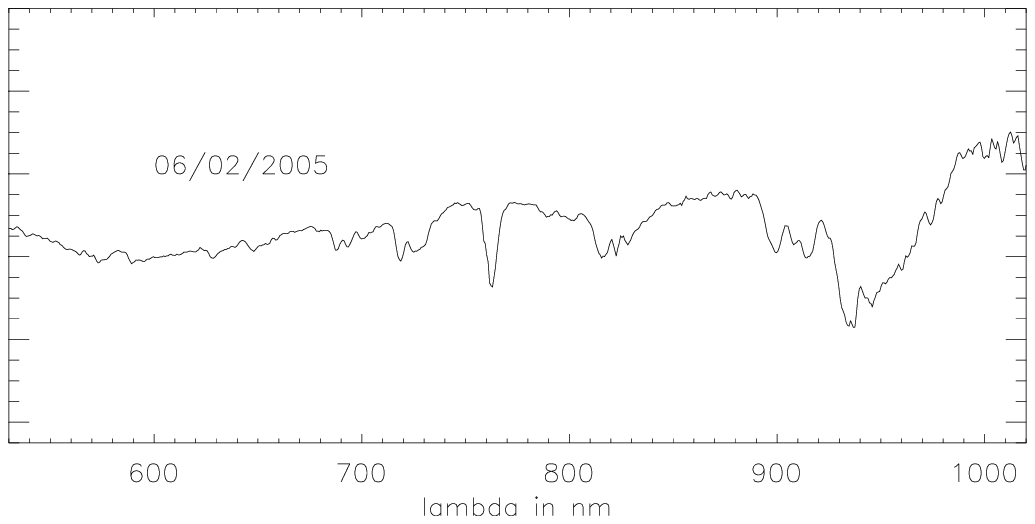}}
\scalebox{0.33}{\includegraphics{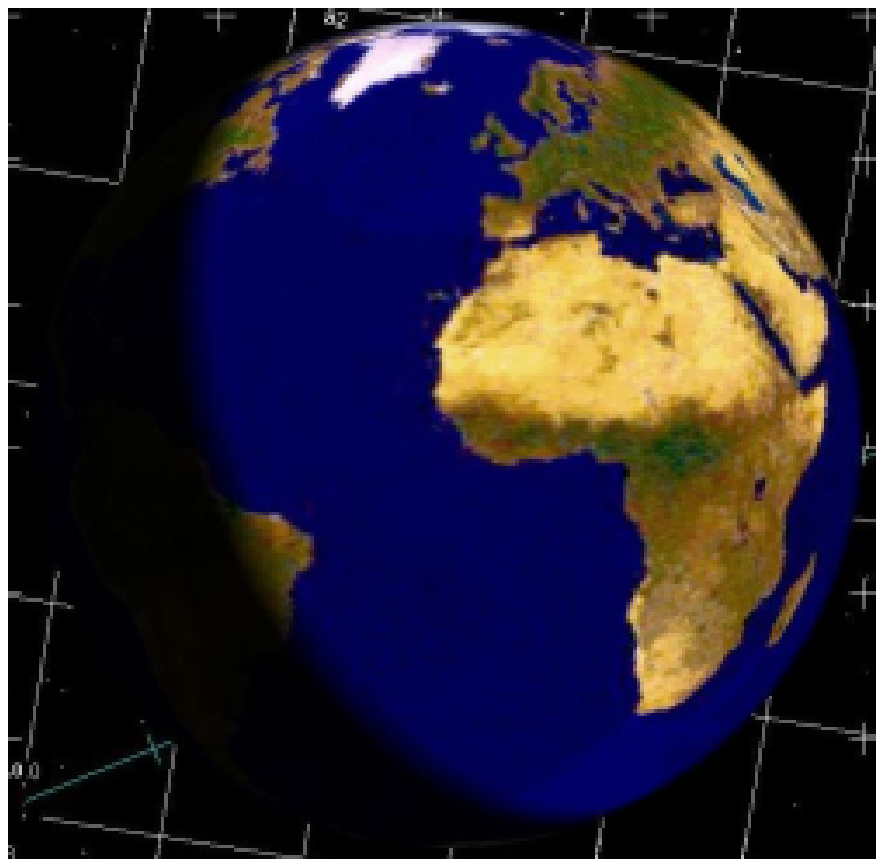}} 

\caption{Earth reflectance spectra. The blue spectra have been scaled to agree (least mean square) with the Rayleigh-aerosol fit of the red spectra. On the right, the Earth seen from the Moon at the epoch of the observation (cloudless model).}
\label{spectres_albedo}
\end{figure*}

\begin{figure}[!]
\scalebox{0.7}{\includegraphics{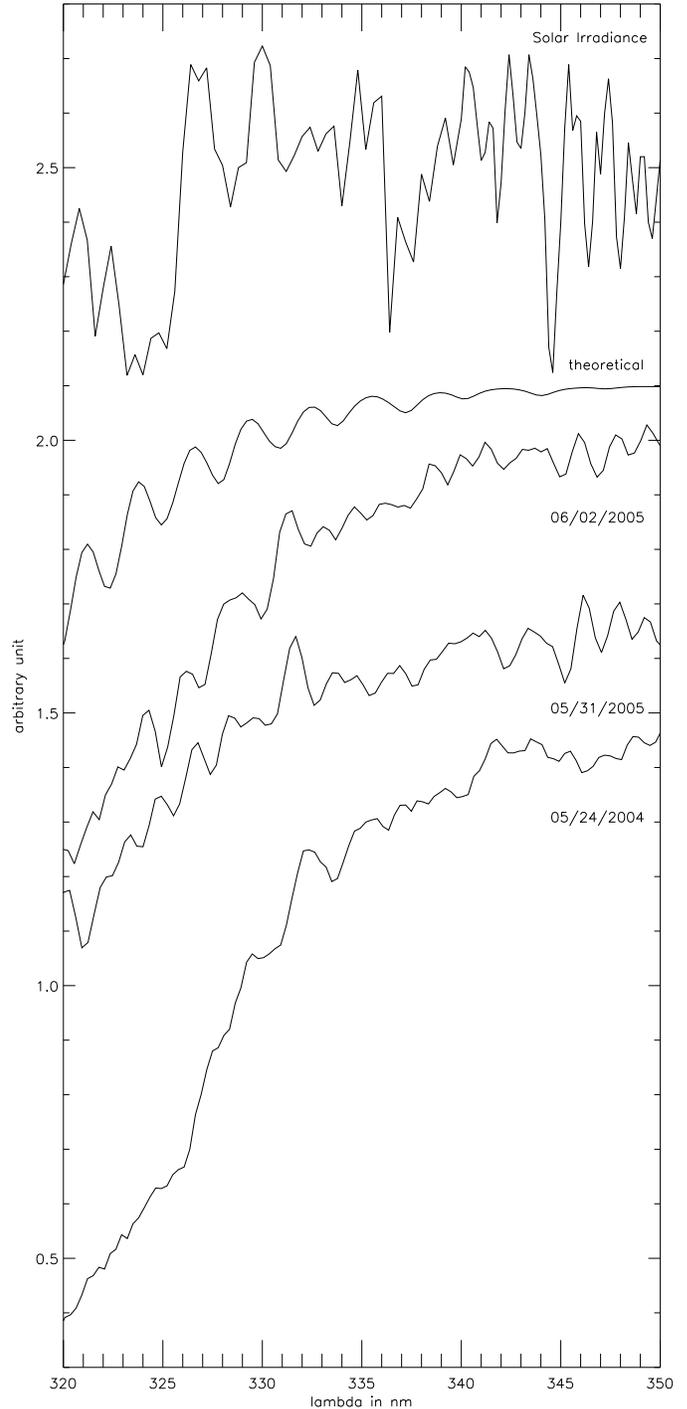}}
\caption{ER($\lambda$) near-UV spectra (three lowest curves) showing the strong absorption of $O_3$ ozone (Huggins band). The theoretical spectrum for ozone is from Voigt et al. (2001). The upper spectrum is the solar irradiance (\cite{arvesen1969}). The noise in the three near-UV ER spectra is due to the low signal in the UV and to imperfect correction of the solar irradiance by the division of Eq.~\ref{albedo}. The spectra are shifted vertically for clarity (+0, +0.3, +0.6, +1.1, +1.6). }
\label{ozone_observation}
\end{figure}

Figures \ref{spectres_albedo} and \ref{ozone_observation} show the obtained ER spectra. The spectra show different features:
\begin{itemize}
\item The blue part (Fig.\ref{spectres_albedo}) displays a decrease above 350 nm due to the Rayleigh scattering. This indicates a partially clear atmosphere and shows that the integrated Earth could be a 'pale blue dot' or a nearly white dot (2nd June 2005 data), given the amount of cloud. 
\item At wavelengths below 350 nm (Fig.\ref{ozone_observation}), reflectance becomes much lower indicating that Earth is clearly a much fainter object in the near UV. This is due to the ozone strong absorption in the UV (Huggins band) giving a high contrast between the visible and UV Earth. The three UV spectra are slightly different due to the different signal to noise ratio of the data. It could be possible to measure the integrated ozone column from the Huggins band (\cite{arnold02}), but we could not investigate this in more details due to the low signal to noise ratio is this spectral domain, cause by the low signal and residual solar spectral features that are not perfectly cancelled out in the division of Eq.~\ref{albedo}. The wider but weaker band of $O_3$ absorption (Chappuis band) peaks at about 600 nm. 
\item Oxygen bands at 630, 690, and mainly at 760 nm are clearly visible. The oxygen dimer ($O_2)_2$ produces a weak absorption band visible between 570 and 580 nm (\cite{tinetti06}), but is superimposed by a weak ozone features from the Chappuis absorption band. The depth of the 760nm band is routinely used in the Earth remote sensing to measure the altitude of clouds by satellites imaging Earth's surface: the deeper the absorption, the lower the altitude of the cloud. Thus in principle the depth of this band in integrated spectra gives an average of the clouds altitude.
On an extra-solar planet, the band's relative depth variation with time would indicate a variation of cloud mean altitude. In our data, the relative depth of $O_2$ shows a larger variation than expected, for a not yet understood reason.
\item Water vapour bands are also clearly visible features, especially in the redder part of the spectrum.
\item Vegetation produces a small increase in the continuum spectrum, barely visible in the 730 to 750 nm range compared to the bluer continuum, and requiring a more quantitative analysis described in the following section. 
\end{itemize}

Since $O_2$ and $O_3$ are product and by-product of photosynthetic life on Earth, the observation of these two molecules in the spectrum of an extra-solar planet could be a potential indicator of life, in combination with other parameters (planet in the habitable zone, i.e. with liquid water).

\subsection{Quantitative analysis}
During the first observing run, both blue and red spectra were not recorded during the same night, but the aim was to record at least a red spectrum of the Earth reflectance showing the vegetation (09/18/2004). The second run allowed us to record all spectra from 320 to 1020 nm in about 40 min.  We measured a VRE between 1.3 and 4.0\% (Table \ref{VRE_data}, Figure \ref{VRE}) with an accuracy of about $\pm0.5$ (absolute error on VRE).

\begin{table}
\caption{Vegetation Red Edge. The $O_3$ concentration is the amount that provides the best (least-square) correction of the Chappuis band. The raw VRE is compute on the reflectance spectra without any atmospheric correction.}
\label{VRE_data}
\centering  
\begin{tabular}{lccc}
\hline\hline
Date  & 2004/09/18  &   2005/05/31 & 2005/06/02 \\
\hline 
\\ 
\textbf{VRE [\%]}  & \textbf{1.3}$\pm0.5$      & \textbf{3.4}$\pm0.5$     & \textbf{4.0}$\pm0.5$   \\
$O_3$ [DU]$\pm3\sigma$ & 1100$\pm600$ & 110$\pm2$ & 740$\pm70$ \\
raw VRE [\%]& 1.4     & 2.0   & 5.8 \\
\end{tabular}
\end{table}

The many steps in the data reduction process required to obtain ER and VRE induce the uncertainties evaluated as follows:

\begin{itemize}
\item Small variations in the width of the continuum used for the determination of the VRE give variations of about $\pm0.1$ (i.e. $VRE=4.0\pm0.1\%$). This comes from the noise in the spectrum over the relatively narrow domains of continuum.
\item The fit of the Rayleigh-aerosol function is based on narrow continuum domains between 520 and 680 nm. Variations in the limits of these domains result in a small $\pm0.05$ on the VRE, but in the noisy spectra of 05-31-2005, the variations can be up to $\pm1.5$. This shows the care needed in the acquisition and processing of the data. 
\item Table \ref{VRE_data} gives the amount of $O_3$ giving the best (least square) Rayleigh-aerosol fit of the data. The mean terrestrial $O_3$ column density is known to be about 300 Dobson Units (DU)\footnote{A Dobson unit (DU) represents the thickness of a layer of an atmospheric gas, here ozone, if it could be compressed to standard ground temperature and pressure. One Dobson Unit is 0.01 mm and the total ozone column is typically 300 DU.}. Considering that the reflectance ER results from two paths through the atmosphere and is integrated on the visible illuminated Earth (see Sect.\ref{reflectance}), we expect to measure an equivalent column of the order of 1200 DU. But we find values from 110 to 1100 DU (Table \ref{VRE_data}), suggesting that the $O_3$ fit based only on the narrow continuum parts of the red side of the ER spectrum is probably not sufficient. A better fit would have been obtained with a full spectrum from 400 to 800 nm to benefit from the full Chappuis band for the fit, but unfortunately it was not possible to overlap both blue and red spectra. The estimation of the $O_3$ concentration results from the minimum root mean square between the fit and the data. The uncertainty in the ozone correction gives a variation of $\pm0.2$ on the VRE.
\item The determination of the synthetic sky background is done by smoothing the variations of the background \textit{slope per ADU}. The width of the smoothing box is determined by the S/N ratio in the ES spectrum. Choosing a different but still reasonable smoothing box produces variations of $\pm0.1$ on the VRE, and if we do not smooth the values, the VRE varies by around $\pm0.5$.
\item The phase-dependant colour of the Moon is estimated from a relatively reduced and sparse number of data points from Lane \& Irvine (1973), but as far as we know, these are the only available at phases above $90^{\circ}$. Comparing different approaches for the extrapolation of the Lane \& Irvine data to larger phase angles, the VRE remains unchanged within $\pm0.1$. But if we do not take into account the phase-colour correction, all VRE values decrease by about 1.5, pointing out the relevance of the colour correction. The residual colour bias of 3\% (end of Sect.\ref{sec:moon}) induces a positive bias in the VRE of $\approx0.15$. To minimize the colour bias, better observations should always observe the same calibrated spots in Earthshine and sunlit Moon to obtain the exact phase-dependent colour and photometry of these areas, as Qiu did for accurate albedo measurements (\cite{qiu03}). 
\end{itemize}

Simple models (\cite{arnold02,arnold03}) predicted a VRE near 8-10\% when Africa and Europe light the Moon. Our results, from 3 to 4\%, are lower, but it is necessary to have more spectra to validate our treatment method or compare with simulations including the real cloud coverage. The result of 1.3\% of VRE for the 09-18-2004, when the Pacific faced the Moon, suggests that there is probably a small positive bias in our VRE estimation method. We also note a small positive slope in the red part of the red spectrum, between 700 and 1000 nm, suggesting the signature of deserts (Sahara). If we correct the spectrum for this increase, all VRE estimations decrease, giving a VRE near 3\% for the Earth showing Africa and Europe. This low VRE may be consistent with the satellite images that show large cloud coverage over rain forests. However, VRE over the Pacific Ocean versus Africa remains different thus allowing the detection of vegetation on Earth.

As described in Section \ref{sec:moon}, the best way to correct for the phase-dependent colour effect of the Moon is to have calibrated photometric data. When we record the Earthshine, we multiplied the reflectance spectra of the Earth by the Moon reflectance and the atmospheric transmittance. The Earthshine is always observed at $0^\circ$ phase. To correct for the Moon reflectance, we thus need a record of the same pattern at Full Moon, corrected from the atmosphere transmittance. To correct for the atmosphere transmission during the observations, we need the spectrum of the Moonshine with a calibrated reflectance for that particular phase.

\section{Conclusion}
We have presented spectra of the Earth reflectance from near-UV to near-IR (320 to 1020 nm) for four different nights. They show significant variations in Rayleigh scattering depending on the cloud cover (the Earth 'blue dot' can be almost white). One of the spectra was taken with mainly light from the Pacific Ocean while the others included parts of or the whole Africa and Europe. The Vegetation Red Edge is observed when land with forests are present, with values between 3 and 4\% ($\pm\approx0.5\%$), but remains very low otherwise ($\approx1\%$). A spectro-photometric accuracy better than 1\% is therefore required for future ground based observations (with Extremely Large Telescopes) or space missions aiming to detect an analog to the terrestrial VRE. These instruments will all be equipped with an optimized coronagraph allowing starlight rejection better than $10^5$ to be able to provide the image of the planet itself. 

A survey over one year or more with monthly observations would allow one to follow the VRE seasonal variations and undoubtedly improve our knowledge of the behavior of this biomarker.  In addition, it would be desirable to have a direct measurement and monitoring from space of the global (i.e. disk-averaged) spectrum of the Earth. One solution would be to observe Moon Earthshine: from space, the Earth reflectance would still be obtained with Eq.\ref{albedo}, but would avoid the source of noise from one path through the atmosphere for both ES and MS light. Other solutions would be to integrate spatially-resolved spectra from low-orbit satellites, or ideally, to have a satellite far enough from the Earth to see it entirely and able to easily obtain disk-averaged spectra of the Earth radiance. The Earth reflectance would then simply be obtained after a division by the solar radiance, avoiding all sources of noise and bias due to the Moon.

Our work showed that measuring the VRE remains very difficult, and, although measurable, the VRE is a small feature when compared to $O_2$, $O_3$ and water vapor absorption lines. But $O_2$ present in the Earth atmosphere is 100\% of photosynthetic origin and its deep absorption line at 760 nm is, for the Earth, a signature of photosynthesis. This suggests that if $O_2$ is observed at 760 nm on an exo-Earth, it is relevant to look for a VRE-like (probably small) feature in the spectrum, keeping in mind that this signature can be significantly different (spectral shape and wavelength) on an exo-Earth than on our planet. Clearly a sharp feature may be much more detectable than a smoother signature. An exhaustive analysis of possible artifacts remains necessary to consolidate the work done on this subject (\cite{schneider04,seager05}). Phylogenetic analysis of plants and research on alternative photosynthesis chemistries are also of prime interest to better constrain this plausible biomarker.


\begin{acknowledgements}
We acknowledge O. Hainaut, P. Leisy and E. Pompei  for their efforts in making these Earthshine observations possible at NTT. 
S. H. was supported by the Swiss National Science Foundation under grant number PBSK2--107619/1.
\end{acknowledgements}


\begin{thebibliography}{}
    \bibitem[Arcichovsky 1912]{arci1912} Arcichovsky V.M.  1912, 'Auf der Suche nach Chlorophyll auf den Planeten' in
				Annales de l'Institut Polytechnique Don Cesarevitch Alexis a Novotcherkassk, Vol. 1,  No 17, 195
    \bibitem[Arnold et al. 2002]{arnold02} Arnold, L., Gillet, S., Lardi{\` e}re, O., Riaud, P. \& Schneider, J. 2002, \aap, 352, 231-237
    \bibitem[Arnold et al. 2003]{arnold03} Arnold, L., Br\'eon, F.M., Brewer, S., Guiot, J., Jacquemoud, S. \& Schneider, J. 2003, in 'SF2A-2003: Scientific Highlights 2003', Bordeaux, France, 16-20 June 2003, eds. F. Combes, D. Barret, T. Contini and L. Pagani, EDP-Sciences, Conference Series, 133-136
    \bibitem[Arvesen et al. 1969]{arvesen1969} Arvesen, J.C., Griffin, R.N. Jr., \& Pearson, B.D. Jr. 1969, Applied Optics, Vol. 8, No 11, 2215-2232.
		\bibitem[Clark 1999]{clark99} Clark R.N. 1999, Manual of Remote Sensing (J. Wiley and Sons, ed. A. Rencz, New-York)
		\bibitem[Danjon 1936]{danjon1936} Danjon A. 1936, Ann. Obs. Strasbourg, 3, 139-180
		\bibitem[Lane \& Irvine 1973]{lane73} Lane, A.~P. \& Irvine, W.~M. 1973, \aj, 78, 267
		\bibitem[L\'ena 1996]{lena96} L\'ena, P., Lebrun, F. \& Mignard, F. 1996, 'M\'ethodes physiques de l'observation', CNRS Editions, EDP Sciences, 2nd edition, 55
		\bibitem[Monta\~ n\' es-Rodriguez et al. 2005]{montanes05} 
				Monta\~ n\' es-Rodriguez, P., Pall\' e, E., Goode, P.~R., Hickey, J. \& Koonin, S.~E. 2005, \apj, 629,1175-1182
		\bibitem[Pall\' e et al. 2004]{Palle04}
				Pall\' e, E., Monta\~ n\'es Rodriguez, P., Goode, P.~R., Qiu, J., Yurchyshyn, V., Hickey, J.,
				Chu, M.-C., Kolbe, E., Brown, C.~T. \& Koonin, S.~E. 2004, Advances in Space Research, 34, 288-292
		\bibitem[Qiu et al. 2003]{qiu03}
    		Qiu, J., Goode, P.~R., Pall\' e, E., Yurchyshyn, V., Hicke, J., Monta\~ n\' es Rodriguez, P., Chu, M.-C.,
				Kolbe, E., Brown, C.~T. \& Koonin, S.~E. 2003, J. of Geophys. Res. (Atmospheres), 108, D22, 12
   	\bibitem[Rougier 1933]{rougier33} Rougier, G. 1933, Annales de l'Observatoire de Strasbourg, 2, 3
   	\bibitem[Russell 1916]{russell16} Russell, H. N. 1916, 	\apj, 44, 128
   	\bibitem[Tinetti et al. 2006]{tinetti06} Tinetti, G., Meadows, V., Crisp, D., Fong, W., Fishbein, E., Turnbull, M. \& Bibring, J.-P. 2006, Astrobiology, 6, 34-47
   	\bibitem[Tikhoff 1914]{tikhoff14} Tikhoff, G.A. 1914, Mitteilungen der Nikolai-Hauptsternwarte zu Pulkowo, no 62, Band $VI_2$, 15
   	\bibitem[Schneider 2004]{schneider04} Schneider, J. 2004, ``Review of visible versus IR characterization of planets and biosignatures'' in 'Towards other Earths: Darwin/TPF and the search for extrasolar terrestrial planets', ESA SP-539, 205
		\bibitem[Seager et al. 2005]{seager05}   		Seager, S., Turner, E.~L., Schafer, J. \& Ford, E.~B. 2005, Astrobiology, 5, 372-390
		\bibitem[Very 1915]{very1915} Very, F.~W. 1915, Astron. Nachrichten, Band 201, Nr. 4819-20, 353-400
		\bibitem[Voigt et al. 2001]{voigt01} Voigt, S., Orphal, J., Bogumil, K. \&d Burrows, J.P. 2001, Journal of Photochemistry and Photobiology A: Chemistry 143, 1-9
		\bibitem[Woolf et al. 2002]{woolf02}    Woolf, N.~J., Smith, P.~S., Traub, W.~A. \& Jucks, K.~W. 2002,		\apj, 574, 430-433

\end{thebibliography}
\end{document}